\documentstyle[12pt]{article}
\def\be{\begin{equation}}
\def\ee{\end{equation}}
\def\ba{\begin{eqnarray}}
\def\ea{\end{eqnarray}}

\def\bq{\begin{quote}}
\def\eq{\end{quote}}
\def\PL{{ \it Phys. Lett.} }
\def\PRL{{\it Phys. Rev. Lett.} }
\def\NP{{\it Nucl. Phys.} }
\def\PR{{\it Phys. Rev.} }

 at 10truept

\newcommand{\beq}{\begin{equation}}
\newcommand{\eeq}{\end{equation}}
\newcommand{\beqa}{\begin{eqnarray}}
\newcommand{\eeqa}{\end{eqnarray}}

\def\ltap{\ \raise.3ex\hbox{$<$\kern-.75em\lower1ex\hbox{$\sim$}}\ }
\def\gtap{\ \raise.3ex\hbox{$>$\kern-.75em\lower1ex\hbox{$\sim$}}\ }
\def\gl{\ \raise.5ex\hbox{$>$}\kern-.8em\lower.5ex\hbox{$<$}\ }
\def\roughly#1{\raise.3ex\hbox{$#1$\kern-.75em\lower1ex\hbox{$\sim$}}}

\begin{document}
\thispagestyle{empty}
\begin{flushright}
SU-ITP-00/04\\ hep-th/0001197\\ January 2000
\end{flushright}
\vspace*{1cm}
\begin{center}
{\Large \bf A SMALL COSMOLOGICAL CONSTANT FROM}

{\Large \bf A LARGE EXTRA DIMENSION}\\
\vspace*{1.2cm}
{\large Nima Arkani-Hamed$^{a,b,}$\footnote{arkani@thsrv.lbl.gov},
Savas Dimopoulos$^{c,}$\footnote{savas@stanford.edu},
Nemanja Kaloper$^{c,}$\footnote{kaloper@epic.stanford.edu}
}\\
\vspace{1.5mm}
{\large and }\\
\vspace{1.5mm}
{\large Raman Sundrum$^{c,}$\footnote{sundrum@leland.stanford.edu}
}\\
\vspace{.8cm}
{\em $^a$
Department of Physics, University of California, Berkeley, CA 94530}\\
\vspace{.15cm}
{\em $^b$Theory Group, Lawrence Berkeley National Laboratory,
Berkeley, CA 94530}\\
\vspace{.15cm}
{\em $^c$
Department of Physics, Stanford University, Stanford, CA 94305}\\
\vspace{1cm}
ABSTRACT
\end{center}
We propose a new approach to the Cosmological Constant Problem which makes
essential use of an extra dimension. A model is presented in which the
Standard Model vacuum energy ``warps'' the higher-dimensional
spacetime while preserving $4D$ flatness. We argue that the strong
curvature region of our solutions may effectively cut off the size of
the extra dimension, thereby giving rise to macroscopic $4D$ gravity
without a cosmological constant.
In our model, the higher-dimensional
gravity dynamics is treated classically with carefully chosen couplings. Our
treatment of the Standard Model is however fully quantum field-theoretic,
and the  $4D$ flatness of our solutions is robust against
Standard Model quantum loops and changes to Standard Model
couplings.

\vfill
\setcounter{page}{0}
\setcounter{footnote}{0}
\newpage

The extremely small value of the cosmological constant
poses the most severe naturalness
problem afflicting fundamental physics (see Ref. \cite{wein} for a review).
The problem stems from the fact that in General Relativity, {\it all} forms of
energy necessarily act as gravitational sources which curve spacetime.
Generically, the energy density of the vacuum as represented by the
cosmological constant, would yield an unacceptably high curvature. An old
but still exciting idea for ameliorating this situation is to assume that
we live in a fundamentally higher-dimensional spacetime, which is indeed
greatly curved by vacuum energy \cite{rusha}.
However, it is now possible that the
curvature spills into the extra dimensions, which
can be larger than the fundamental Planck length \cite{add,aadd,rso,rst,addk1}.
This idea has never been realized in any way
which does not require the same level of fine-tuning as in
four dimensions, although the way the fine-tuning shows up does change in
interesting ways \cite{rso,rst,comp,nk,dwfgk,Myers}. A very interesting
recent approach to the problem has been explored in
\cite{verlindeold,verlinde,schmid}.

In this paper, we report on some limited progress in this direction.
Working within  a five-dimensional
``brane universe'' effective field theory \cite{sun} in which
the Standard Model (SM) is confined to a 3-brane, we will
show that one can carefully
choose the  higher-dimensional gravitational dynamics so that
the SM vacuum energy is effectively converted into a current
which is carried off the brane into the ``bulk''.
The gravitational back-reaction of this
current warps the higher-dimensional spacetime, but in a manner consistent
with four-dimensional Poincare invariance of the vacuum solution and
inconsistent with four-dimensional de-Sitter or Anti-de-Sitter symmetry.
However, our vacuum solution necessarily contains a singular region
parallel to the SM brane, signaling high sensitivity to short-distance
gravity and the breakdown of effective field theory in that region.
Nevertheless we will argue that the
singular region may effectively cut off the higher dimensional spacetime
as in \cite{gmz,CK} (see also \cite{brsf}),
thereby rendering the gravitational dynamics macroscopically
four-dimensional, with vanishing
effective cosmological constant.

In this paper we will explicitly consider only
classical (higher-dimensional) gravity, but with  quantum SM effects
included. The gravitational couplings must be chosen very precisely for the
mechanism to work, although having chosen them the mechanism is
completely stable under SM radiative corrections and changes in SM
couplings. To be precise about the size of effects neglected by the
truncation to classical gravity, it is useful to
distinguish the contributions
to the cosmological constant which do and do not involve quantum gravity,
whose fundamental scale is $M_*$.
In an effective field theory framework valid up to an energy
$\Lambda < M_{*}$, we can expand the cosmological constant as
\begin{equation}
\lambda = {\cal O}(\Lambda^4) + {\cal O}(\frac{\Lambda^6}{M_{*}^2}) +
...~ .
\label{approx}
\end{equation}
The first term obtains a contribution from the SM vacuum energy,
treated fully quantum field theoretically, but is insensitive to quantum
gravity corrections. The subsequent terms are sensitive to quantum
gravity effects, described in Feynman diagrams by virtual graviton
exchanges  suppressed by powers of $1/M_{*}^2$. Note that
the formally  larger first term is under the best theoretical control,
while the subleading terms are shrouded to some extent by the mysteries of
quantum gravity. Our truncation limits us to understanding the
screening of the ${\cal O}(\Lambda^4)$ SM vacuum energy, deferring
the problem of ${\cal O}(\frac{\Lambda^6}{M_{*}^2})$
effects. This
does not mean that these subleading effects are
phenomenologically negligible
since the smallest we can take $\Lambda$ is our experimental cutoff of
${\cal O}$(TeV), so that even  $\frac{\Lambda^6}{M_{*}^2}$ is much
larger than the observed cosmological constant.
Nevertheless we consider it a significant
advance to  show how the effects of the SM vacuum energy
alone can  be nullified.

Our set-up is as follows. We consider a five-dimensional spacetime, or
``bulk", where the fifth dimension  is a half-line with
coordinate $y$, while the usual
four dimensions have coordinates $x^{\mu}$. Orbifold boundary conditions
as in \cite{losw,rso} will be used to realize the half-line
in terms of  a full real line with the identification of $y$ with $-y$.
The five-dimensional degrees of freedom will
therefore be taken to be symmetric about $y = 0$.  We will consider the
boundary of the half-line, $y=0$, to be the location of an
``end-of-the-world'' 3-brane, to which the SM quantum field theory
(or some extension thereof) is confined. The SM degrees of freedom
interact with classical five-dimensional bulk gravity as well as a
classical five-dimensional scalar field $\phi$.
Similar models have been studied in different contexts
in \cite{losw,gw,dwfgk}.
In the bulk, the scalar field will only be coupled minimally to
bulk gravity, and on the brane it will be conformally coupled to the SM.
Similar couplings have been employed in $4D$ attempts to solve
the cosmological constant problem \cite{conf,wein}.
We will carefully choose the conformal coupling constant such that
for a given value of the vacuum energy on the brane there
exists a flat solution.
However, as we will show, once this choice is made, the
ensuing flat solution remains unaffected by quantum loops
on the brane, showing that the vanishing of the $4D$ cosmological
constant is both radiatively stable and insensitive to the
values of the SM parameters. Up to two derivatives,
the action of our model is
\be
S =  \int d^4x dy \sqrt{g_5} \Bigl(\frac{R}{2 \kappa^2_5}
- \frac32 (\nabla \phi)^2 \Bigr)
-\int d^4 x
\sqrt{-\det(g_4 (0) e^{\kappa_5 \phi(0)})}
{\cal L}_{SM}(H, g_{\mu\nu}(0) e^{\kappa_5 \phi(0)}) \, .
\label{action}
\ee
The constant $\kappa^2_5$
is related to the $5D$ Planck scale $M_*$ by $\kappa^2_5 =
M^{-3}_*$. Here $M,N$ run
over the $5D$ coordinates, while $\mu,\nu$ run over the
$4D$ brane coordinates.
Note the special normalization of the bulk scalar $\phi$,
or equivalently the special coupling to the brane for the
canonically normalized scalar field.
The SM Lagrangian ${\cal L}_{SM}$
is localized to the brane at $y=0$. It depends on the SM fields,
$H(x)$ and any necessary ultraviolet regulators, all
minimally coupled to the Weyl-rescaled induced metric
$g_{\mu\nu}(y=0) e^{\kappa_5 \phi(y=0)}$.
Note that we have set the bulk cosmological constant and the
operators $f(\kappa_5 \phi) R$ equal to
zero. With the Weyl transformation
$\bar g_{MN} = g_{MN} \exp(\kappa_5
\phi)$ we can cast the action (\ref{action}) into the
``string frame", where it takes the form
\be
S =  \int d^4x dy \sqrt{\bar g_5} e^{-3\kappa_5 \phi/2}
\frac{\bar R}{2\kappa^2_5} - \int d^4 x
\sqrt{\bar g_4(0)}
{\cal L}_{SM}(H, \bar g_{\mu\nu}(0)) \, .
\label{saction}
\ee
It is now clear from this action that our choice of the
conformal coupling is completely unaffected by quantum
loops on the brane. The brane action, (including ultraviolet regulators),
does not even contain the scalar field $\phi$, so $4D$ general
covariance guarantees that the SM renormalization
will not change the form of the actions (\ref{saction}) and (\ref{action}). For
simplicity, we will use the canonical action (\ref{action})
in what follows.

To incorporate the effects of SM quantum loops we can
integrate out the SM degrees of freedom, and work in terms of
the full 1PI effective action. Hence
the brane action in (\ref{action})
is replaced by the effective action,
\be
S_{brane} \rightarrow
\Gamma^{SM}_{eff}(H,g_{\mu\nu}(0)
e^{\kappa_5 \phi(0)})\, .
\label{effact}
\ee
The equations of motion are now straightforward to obtain.
Varying the action (\ref{action}) and using (\ref{effact})
we find
\ba
R^{MN} - \frac12 g^{MN} R &=& 3 \kappa^2_5
\Bigl(\nabla^M \phi \nabla^N \phi - \frac12 g^{MN}
(\nabla \phi)^2 \Bigr) +
\frac{2 \kappa^2_5}{\sqrt{g_5}}
\frac{\delta \Gamma^{SM}_{eff}}{\delta g_{\mu\nu}}
\delta^M{}_\mu \delta^N{}_\nu \delta(y) \, ,
\label{eombone} \\
3\Box_5 \phi &=& - \frac{1}{\sqrt{g_4}}
\frac{\delta \Gamma^{SM}_{eff}}{\delta \phi} \delta(y) \, ,
\label{eombtwo} \\
\frac{\delta \Gamma^{SM}_{eff}}{\delta H} &=& 0 \, .
\label{eombthree}
\ea
Equation (\ref{eombthree}) encapsulates the full SM
quantum field theory in the curved background.

We now seek solutions with $4D$ Poincare symmetry.
With this symmetry the SM effective action is
given by just the effective potential,
\be
\Gamma^{SM}_{eff} = -\int d^4x \sqrt{g_4}
V_{eff}(H) e^{2\kappa_5\phi}\, ,
\label{effpot}
\ee
and hence the
SM equations of motion take the simple form
\be
\frac{\partial V_{eff}}{\partial H} = 0 \, .
\label{extre}
\ee
We will denote an extremal value of the potential by $V_{extremal}$.
Further, the structure of the $5D$ metric is restricted by the
Poincare symmetry to the form
\be
ds^2 = a^2(y) \eta_{\mu\nu} dx^\mu dx^\nu + dy^2 \, ,
\label{metans}
\ee
where $a(y)$ is the warp factor. The equations of motion
(\ref{eombone})-(\ref{eombtwo})
then take a particularly simple form.
Defining the shifted scalar field $\tilde \phi =
\phi + \frac{1}{2\kappa_5} \ln(\frac{V_{extremal}}{M^4_*})$,
the field equations in the bulk become
\be
\frac{a'^2}{a^2} = \frac{\kappa^2_5 \tilde \phi'^2}{4}\, ,
~~~~~~~~~~~
\tilde \phi'' + 4 \frac{a'}{a}\tilde \phi' = 0 \, ,
~~~~~~~~~~~
\frac{a''}{a} = - \frac{3\kappa^2_5 \tilde \phi'^2}{4} \, ,
\label{eqsb}
\ee
while the $\delta$-function sources and the $y \rightarrow -y$
symmetry imply the matching conditions for the first derivatives
\be
a'(0) = - \frac{M_*}{6} e^{2\kappa_5 \tilde \phi(0)} a(0)\, ,
~~~~~~~~~~~ \tilde \phi'(0) = \frac{M^{5/2}_*}{3}
e^{2\kappa_5 \tilde \phi(0)} \, .
\label{bcs}
\ee
Here the prime refers to the derivative with respect to $y$.
Note that with the definition of the variable
$\tilde \phi$, the gravitational field equations
are completely independent of the extremum value of the
SM effective potential $V_{extremal}$. This will be crucial
for ensuring the success of our mechanism.

It is straightforward to solve the system
(\ref{eqsb})-(\ref{bcs}).
The explicit solutions with Poincare symmetry are
\ba
ds^2_5 &=& (1- \frac{2M_*}{3}
e^{2 \kappa_5 \tilde \phi_0} |y|)^{1/2}
\eta_{\mu\nu} dx^\mu dx^\nu + dy^2 \, , \nonumber \\
\tilde \phi &=& \tilde \phi_0 -\frac{1}{2\kappa_5}
\ln\Bigl( 1- \frac{2 M_*}{3}
e^{2 \kappa_5 \tilde \phi_0} |y|\Bigr) \, , \nonumber \\
\phi &=& \phi_0
-\frac{1}{2\kappa_5}
\ln\Bigl( 1- \frac{2 V_{extremal}}{3 M^3_*}
e^{2 \kappa_5 \phi_0} |y|\Bigr) \, ,
\label{soln}
\ea
where $\phi_0$ is an integration constant.
We have checked explicitly that the solutions
with $4D$ Poincare symmetry are the only allowed
solutions which are $4D$ maximally symmetric.
Making a more general ansatz for the metric with the
symmetries of $4D$ (Anti-) de-Sitter space with curvature
$h$, it is straightforward to show that if the scalar field
coupling to the brane is $\exp(\zeta\kappa_5 \phi)$,
the $4D$ curvature is
\be
h^2 = a'^2(0)
- \frac{\kappa^2_5 \tilde \phi'^2(0)}{4} a^2(0)\, ,
\label{fdcurv}
\ee
which vanishes by virtue of our choice of the conformal
coupling coefficient $\zeta=2$ in eq. (\ref{action}).
This coupling {\it prohibits} both de-Sitter
symmetry and Anti-de-Sitter symmetry
on the brane, regardless of the details of the SM
physics. The solutions (\ref{soln}) give a flat Minkowski space, for any
value of $V_{extremal}$ even after all quantum corrections are
included. Further, altering the parameters of SM, such as for
example the electron mass and the fine structure constant, will
not break the Poincare symmetry.

The fact that we have found a solution with $4D$ Poincare invariance,
robust against SM loops and couplings, does not complete our task.
For instance, we can
always find $4D$ flat solutions for a 3-brane
carrying a range of tensions in $6D$, the tension
just inducing a deficit angle in the bulk. The problem is that gravity
remains six-dimensional at long distances.
Therefore, we must ask whether our $5D$
set-up leads to $4D$ gravity at long distances. If it does, then our
Poincare invariant solution demonstrates that the effective $4D$
gravitational dynamics has vanishing cosmological constant.

An important feature of our solution which gives hope for ensuring
macroscopic $4D$ gravity is the appearance of a naked curvature
singularity at finite proper distance,
\be
y_s = \frac{3}{2 M_*} e^{-2\kappa_5 \tilde \phi_0} =
\frac{3 M^3_*}{2 V_{extremal}} e^{-2\kappa_5 \phi_0},
\label{singdist}
\ee
away from the brane. Similar singularities have appeared in the
work of \cite{gmz,CK}.
The appearance of our singularity can be understood  by noting the
remarkable analogy \cite{verlinde}
between our equations and Einstein's equations
in cosmological Friedmann-Robertson-Walker (FRW) spacetimes
(with four spatial dimensions),
if we interpret the coordinate $y$ as a cosmological ``time''
and the warp factor as the FRW ``scale factor''.
The bulk equations (\ref{eqsb}) then coincide
with the cosmological equations of the
FRW universe dominated by a massless scalar field.
This should not be a surprise, since the ansatz of (\ref{metans})
is a Wick rotation of the FRW ansatz.
The analogy with cosmological dynamics
immediately shows that away from the brane
only two possibilities can occur. One possibility is that the warp factor
$a$ monotonically increases  forever and the scalar
field dissipates away, which means
that the topology of the extra dimension is not compact
and manifestly has infinite volume. This case however
requires negative energy on the brane. Our solution
follows from $V_{extremal}>0$,
namely the warp factor monotonically decreases
to zero at a finite distance from the
brane, where the curvature and the scalar field diverge.
This FRW analogy can be thought of in two ways related by time reversal:
either the brane specifies
initial conditions which evolve into a ``Big Crunch", or a
``Big-Bang" evolved with final conditions specified by the brane.

Clearly our solution cannot be trusted for $y > y_s$. A careful
discussion of the singularity becomes crucial in deciding whether
gravity becomes four-dimensional at distances larger than
$y_s$.
That we need to do this at all is already
striking: the singularity of our solution forces the {\it short-distance}
properties of quantum gravity to become relevant to whether we recover the
inverse square law for gravity {\it at long distances}!
We will assume that the singularity is smoothed out by the true short-distance
theory of gravity. Since we do not know the details of this theory, we will
not be able to make any rigorous claims about whether the resolution of the
singularity does what we want. Nevertheless, we see that
even without a detailed
understanding of the physics which smooths out the singularity, there are two
qualitatively different possibilities.
The first is an analog of a ``Big Crunch/Big Bang'' transition.
That is, the theory extends across the singularity
into a region where a weakly coupled Einstein gravity description
is valid again, and the warp factor $a(y)$ diverges as $y \to \infty$.
In this case the singularity {\it does not} end space, and clearly there
is no $4D$ gravity at long distances.
The second and more attractive possibility is that space ends with a
finite volume at
the singularity (the analog of time starting at the Big Bang).
For instance, short-distance effects might smoothly join the region of
high curvature to a highly curved $AdS_5$ geometry;
pushing the singularity to infinite proper distance
while preserving a finite volume space. In this case, the extra dimension is
effectively compactified in the manner of \cite{rst},
and we should get $4D$ gravity at long distances.
Alternatively, any form of spacetime description might become
impossible beyond $y_s$. In either of these cases,  the
zero-mode tensor fluctuations
\ba
ds^2_5 &=& (1- \frac{2M_*}{3}
e^{2 \kappa_5 \tilde \phi_0} |y|)^{1/2}
\overline{g}_{\mu\nu}(x) dx^\mu dx^\nu + dy^2,
\label{4graviton}
\ea
would correspond to a massless $4D$ graviton, with a {\it finite}
 $4D$ Planck scale,
\be
M^2_{Pl} = M_*^3 \int_0^{y_s} a^2(y) = M^2_* e^{-2\kappa_5 \tilde
\phi_0} = \frac{M^6_*}{V_{extremal}} e^{-2\kappa_5
\phi_0} \, .
\label{fdpl}
\ee
Similarly the integration constant $\tilde{\phi}_0$ is promoted to a
$4D$ scalar zero-mode $\tilde{\phi}_0(x)$\footnote{Of course,
a massless, gravitationally coupled scalar field is experimentally
excluded. In a realistic theory, this scalar will have to pick up
a mass of at least $\sim$(mm)$^{-1}$.}.

Summarizing: we have found a solution with a 3-brane in $5D$,
interacting with bulk gravity and a massless scalar.
The bulk couplings are carefully chosen to guarantee $4D$ flat
space solutions,
while prohibiting other maximally symmetric solutions with
$4D$ (Anti-) de Sitter symmetry.
The solutions have a curvature singularity at a finite
distance from the brane, analogous to the
Big Bang or Big Crunch singularity of FRW cosmology. Just as it is
possible to think of the crunch singularity as ``ending time",
it is natural to assume that our singularity ends space,
effectively compactifying the extra dimension and giving $4D$
gravity at long distances. {\it If this assumption is correct},
our model represents a partial solution to the Cosmological Constant
Problem, where SM fine-tuning is avoided.
Note that as the value of the SM vacuum energy, the warp factor
adjusts itself to maintain the vanishing of the $4D$ cosmological
constant. This is similar in spirit to the ideas of
\cite{verlindeold,verlinde,schmid}. Unlike the proposal of
\cite{verlindeold,verlinde,schmid}, in our model the SM fields
reside on the ``Planck" brane, rather than in the bulk.

But how can we be sure that the same short-distance physics which resolves the
singularity does not upset our mechanism ensuring the
appearance of a Poincare invariant solution, and hence vanishing $4D$
cosmological constant, without SM fine-tuning? A way of
``resolving'' the singularity which does not work
is by simply ending space with a second brane.
For instance, a standard
orbifold compactification with a second brane cutting off the singular
region with negative energy $V_{neg}$ does require the precise
fine-tuning, $-V_{neg}=V_{extremal}$ \cite{losw,rso}.
Alternatively one can  note that
our bulk Lagrangian, with the special scalar
couplings, arises precisely from compactifying one of the dimensions
of a $6D$ theory, with the  3-brane coupling only to the induced metric,
and $\phi$ being the modulus of the extra dimension.
It is straightforward to lift our $5D$ solution into the full $6D$
theory. The $6D$ space is compact and
locally flat away from our brane, except at a conical
singularity corresponding to our singularity at $y_s$.
If this conical singularity is described from the outset as a
brane in the theory, its tension must again be fine-tuned against the SM
vacuum energy.

We will formulate the conditions for the short distance physics
under which our mechanism works.
First, we assume
that in the absence of the SM 3-brane the full bulk dynamics admits $4D$
Poincare invariant
solutions describing $5D$ spacetime
ending in the singularity region.
We will focus on these because they are the
only ones compatible with the matching conditions on the brane.
In the analog FRW picture, this amounts to saying that any
spatially flat universe can end in a Big Crunch \cite{verlinde}.
We also assume that the values of $\phi$ and $\phi'$ emerging
from the brane are
always compatible with the physics of the singularity. In the FRW
analogy, this corresponds to saying that a few Planck times after
the Big Bang, a scalar field without a potential can emerge with
some $\dot {\phi}$, but with any $\phi$.
The resolutions of the singularity discussed above failed because
they violated these requirements.
A sufficient but not necessary condition for the second assumption
to be valid is if the short distance physics resolving the singularity
is shift symmetric under $\phi \to \phi +$ constant.
Also note that the contribution to the $4D$ cosmological constant
from higher derivative bulk operators
is Planck-suppressed relative to
$V_{extremal}$ and therefore of the order we are already neglecting
in this paper, as discussed in Eq. (\ref{approx}).

A physical picture of our mechanism follows from the observation
that  shift symmetry of the  bulk action implies
an associated conserved current,
\begin{equation}
J^M = - \frac{1}{\sqrt{g_5}} \frac{\delta S_{bulk}}{\delta \partial_M \phi}.
\end{equation}
$4D$ Poincare invariance implies that only $J_y$ can be non-zero.
The fact that the $\phi$-couplings to the SM on the brane explicitly break
shift symmetry corresponds to a brane-localized source for $J$,
thereby determining the current to be
\begin{equation}
J_y = 3 \phi' = M_*^{5/2} e^{2\kappa_5 \tilde \phi} =
\frac{V_{extremal}}{M_*^{3/2}} e^{2\kappa_5 \phi}.
\end{equation}
Thus we see that in our set-up the SM vacuum energy is converted into a
current emerging from the 3-brane and ending in the singularity region.
While the vacuum energy does not show
up as $4D$ curvature, the gravitational backreaction of the associated
current warps $5D$ spacetime and gives rise to the singularity into which
it pours. As discussed, the singularity might be resolved into a highly
curved $AdS_5$ geometry so that the current has room to continue
infinitely far away from the brane, or possibly spacetime really ends at the
singularity, in which case there may be some non-perturbative breakdown of
shift symmetry which allows the current to end.

We conclude by making some comments on the long-distance 
$4D$ effective theory
in our set-up. While we do not know the details of short-distance gravity
and are forced to speculate on the nature of the singularity,
the physics must yield a consistent $4D$ effective field theory.
In particular, this theory must reflect the property of our mechanism
that all the extrema of $V(H)$ lead to $4D$ flat solutions.
Therefore, the naive guess that the effective theory simply contains
the term
\be
\int d^4x \sqrt{g_4} \Bigl( V_{eff}(H) - V_{extremal} \Bigr)
\label{lagr}
\ee
cannot work because of the possibility of multiple extrema $V(H_1),
V(H_2), ...$. If we subtract the extremum from one vacuum, any
other extremum would gravitate.
If $H$ is the only light field in the theory, the Lagrangian (\ref{lagr})
is the only one consistent with the flat space limit.
However, the presence of other light gravitationally coupled fields
$\psi$ offers a way out.
For example, the true effective $4D$ potential may have the form
$V_{eff}(H,\psi) = F(\psi) V_{4}(H) - G(\psi)$. In the limit
where $M_{Pl} \to \infty$, the $\psi$ decouple and the $H$ dynamics
is governed by $V_{eff}(H)$.
However, with finite $M_{Pl}$, it is possible for
$V_4(H,\psi)$ to only have extrema at points $(H_1,\psi_1),
(H_2,\psi_2)$ with vanishing potential, while
$(H_2,\psi_1)$, $(H_1,\psi_2)$ are not extrema because they
do not satisfy the $\psi$ equation of motion. It is
straightforward to choose functions $F$ and $G$ with this
property.
In our model a natural candidate for the modulus is $\phi$; furthermore,
such a potential between $\phi$ and the electroweak symmetry breaking sector
can naturally induce a mass for $\phi$ of order
(TeV$)^2/M_{Pl} \sim$(mm$)^{-1}$
as required phenomenologically.

In this paper, we have presented a model with a $3$-brane in five
dimensions, whose only maximally symmetric solutions are $4D$
Poincare invariant, independent of the SM parameters.
Our solutions are forced into a strong
curvature region which connects the fate of long distance gravity
with its short distance properties. With specific
assumptions on the nature of the singularity, we recover
macroscopic $4D$ gravity with vanishing cosmological constant,
in a manner consistent with a $4D$ effective field theory.
A better understanding of the singularity
would allow us to establish
or exclude this idea.

\vspace{.5cm}
{\bf Acknowledgements}

We would like to thank Andy Cohen, Michael Dine, Bob Holdom,
Joe Polchinski, Lisa Randall and Herman Verlinde for discussions.
The work of N.A-H. has been supported in part by the DOE under Contract
DE-AC03-76SF00098, and in part by NSF grant PHY-95-14797.
The work of S.D., N.K. and R.S. has
been supported in part by NSF Grant PHY-9870115.

\vskip.3cm
{ Note added:
While we were completing this paper, we were informed of the upcoming work
\cite{kms} which overlaps with the ideas presented here.}


\begin{thebibliography}{99}

\bibitem{wein}
S. Weinberg, {\it Rev. Mod. Phys.} {\bf 61}, 1 (1989).

\bibitem{rusha}
V.A. Rubakov and M.E. Shaposhnikov, \PL {\bf B125}, 136 (1983);
\PL {\bf B125}, 139 (1983).

\bibitem{add} N. Arkani-Hamed, S. Dimopoulos and G. Dvali,
{\it Phys. Lett.} {\bf B429}, 263 (1998);
\PR {\bf D59}, 086004 (1999).

\bibitem{aadd}
I. Antoniadis, N. Arkani-Hamed, S. Dimopoulos and G. Dvali,
{\it Phys. Lett.} {\bf B436}, 257 (1998).

\bibitem{rso}
L. Randall and R. Sundrum, \PRL {\bf 83}, 3370 (1999).

\bibitem{rst}
L. Randall and R. Sundrum, \PRL {\bf 83}, 4690 (1999).

\bibitem{addk1}
N. Arkani-Hamed, S. Dimopoulos, G. Dvali
and N. Kaloper, \PRL {\bf 84}, 586 (2000);\\
J. Lykken and L. Randall, hep-th/9908076.

\bibitem{comp}
R. Sundrum, \PR {\bf D59}, 085010
(1999);\\
N. Arkani-Hamed, S. Dimopoulos, J. March-Russell,
hep-th/9809124.

\bibitem{nk}
N. Kaloper, \PR {\bf D60}, 123506 (1999).

\bibitem{dwfgk}
O. DeWolfe, D.Z. Freedman, S.S. Gubser and A. Karch,
hep-th/9909134.

\bibitem{Myers}
C.P. Burgess, R.C. Myers and F. Quevedo, hep-th/9911164.

\bibitem{verlindeold}
H. Verlinde, hep-th/9906182.

\bibitem{verlinde}
J. de Boer, E. Verlinde and H. Verlinde, hep-th/9912012;\\
E. Verlinde and H. Verlinde, hep-th/9912018.

\bibitem{schmid}
C. Schmidhuber, hep-th/9912156.

\bibitem{sun}
R. Sundrum, \PR {\bf D59}, 085009 (1999).

\bibitem{gmz}
M. Gell-Mann and B. Zwiebach, \NP {\bf B260}, 569 (1985).

\bibitem{CK}
A.G. Cohen and D.B. Kaplan, hep-th/9910132.

\bibitem{brsf}
L. Girardello, M. Petrini, M. Porrati and A. Zaffaroni,
{\it JHEP}  {\bf 9905}, 026 (1999);\\
A. Brandhuber and K. Sfetsos, {\it JHEP} {\bf 9910}, 013 (1999).

\bibitem{losw} A. Lukas, B.A. Ovrut, K.S. Stelle and D. Waldram,
\PR {\bf D59}, 086001 (1999).

\bibitem{gw}
W.D. Goldberger and M.B. Wise, \PRL {\bf 83}, 4922 (1999).

\bibitem{conf} A.D. Dolgov, {\it The very early universe; Proc.
1982 Nuffield Workshop at Cambridge}, ed. G.W. Gibbons, S.W.
Hawking and S.T.C. Siklos, (Cambridge University);\\
F. Wilczek,
{\it Phys. Rep.} {\bf 104}, 103 (1984);\\
R. D. Peccei, J. Sola and C.
Wetterich, \PL {\bf B195}, 183 (1987);\\
S. M. Barr and D.
Hochberg,  \PL {\bf B211}, 49 (1988).

\bibitem{kms} S. Kachru, M. Schulz and E. Silverstein, to appear.

\end{thebibliography}
\end{document}